\documentclass[aps,preprint,amsmath,amssymb]{revtex4}

\usepackage{graphicx}
\usepackage{color}

\begin{document}
\title{\Large \bf Production of doubly charmed baryons in $B$ decays }
\date{\today}
\author{\large \bf
Chuan-Hung~Chen$^{1,2}$\footnote{Email: physchen@mail.ncku.edu.tw}
 }
\affiliation{
$^{1}$Department of Physics, National Cheng-Kung University, Tainan, 701 Taiwan \\
$^{2}$National Center for Theoretical Sciences, Taiwan }

\begin{abstract}
We study the doubly charmed baryonic $B$ decays $B\to \Xi_{c}
\Lambda_{c}$ and $B\to \Lambda^{+}_{c} \Lambda^{-}_{c} K$, recently
observed by BELLE. We find that the unexpected large branching
ratios (BRs) could be ascribed to the final state interactions
(FSIs), which are dictated by $B\to DD^{+}_{s}\to \Xi_{c}
\Lambda^{+}_{c}$ and $B\to \bar D^0 D^{0} K\to \Lambda^{+}_{c}
\Lambda^{-}_{c} K$. By utilizing the same mechanism, we predict that
the BRs for $B^{+}\to \bar \Xi^{0}_{c} \Sigma^{+}_{c}$ and $B^{+}
\to \Sigma^{\mp}_{c} \Lambda^{\pm}_{c} K^{+}$ decays could be as
large as $BR(B^{+}\to \bar\Xi^{0}_{c} \Lambda^{+}_{c})$ and
$10^{-4}$, respectively. In addition, extending the FSIs to the
processes associated with the creation of the $s\bar s$ pair,
$BR(B^{+}\to \bar \Omega_{c} \Xi^{+}_{c})$  at percent level is
achievable.

\end{abstract}
\maketitle

There have been many baryonic $B$ decays studied at $B$ factories.
They are the charmless decays $B\to p(\Lambda) \bar p(\bar \Lambda)
K$, $B\to p \bar{\Lambda} \pi$ \cite{charmless_baryon1} and $B\to
p(\Lambda) \bar p(\bar \Lambda)$ \cite{charmless_baryon2}, in which
the three-body baryonic decays have branching ratios (BRs) in
magnitude of $O(10^{-6})$ while the BRs of two-body decays are
limited to be less than $6.9\times 10^{-7}$. In addition, the single
charmed baryonic decay $B^{0}\to \bar \Sigma^{--}_{c} p \pi^{+}$ is
measured to be $(2.38^{+0.63}_{ -0.55}\pm 0.41 \pm 0.62)\times
10^{-4}$ \cite{charm_baryon1} while $\bar B^{0}\to \Lambda^{+}_{c}
\bar p $ and $B^{-} \to \Sigma_{c}(2455)^{0} \bar p$ are
$(2.19^{+0.56}_{-0.49}\pm 0.32 \pm 0.57) \times 10^{-5}$ and
$(3.67^{+0.74}_{-0.66}\pm 0.36 \pm 0.95)\times 10^{-5}$
\cite{charm_baryon2}, respectively. Phenomenogically, the BRs of
three-body baryonic $B$ decays are roughly one order of magnitude
larger than those of corresponding two-body decays. The preference
could be understood by the threshold enhancements occurring in the
near threshold of baryon-pair invariant mass \cite{HS,CCT}.
Moreover, one can easily find that in both three-body and two-body
decays, the BRs for single charmed baryonic processes are two orders
of magnitude larger than those for charmless decays. The differences
could be attributed to the Cabibbo-Kobayashi-Maskawa (CKM)
\cite{CKM} matrix elements, where single charmed (charmless)
processes are associated with $V_{cb}(V_{ub})$.  A more detailed
review could be referred to Ref. \cite{Cheng}.

However, when BELLE observes the two charmed baryons in the final
state, the characters appearing in the single charmed and charmless
decays subsequently are changed. According to BELLE's results, the
BRs $BR(B^{+}\to \Lambda^{+}_{c} \Lambda^{-}_{c}
K^{+})=(6.5^{+1.0}_{-0.9}\pm 1.1 \pm 3.4)\times 10^{-4}$ and $
BR(B^{0}\to \Lambda^{+}_{c} \Lambda^{-}_{c}
K^{0})=(7.9^{+2.9}_{-2.3}\pm 1.2\pm 4.1)\times 10^{-4} $ are
measured with statistical significance of $16.4\sigma$ and
$6.6\sigma$,  and the products of BRs for two-body decays are
$BR(B^{+} \to \bar \Xi^{0}_{c} \Lambda^{+}_{c})\times BR(\bar
\Xi^{0}_{c}\to \bar \Xi^{+} \pi^{-})=(4.8^{+1.0}_{-0.9}\pm 1.1 \pm
1.2)\times 10^{-5}$ and $BR(B^{0} \to \bar \Xi^{-}_{c}
\Lambda^{+}_{c})\times BR(\bar \Xi^{-}_{c}\to \bar \Xi^{+} \pi^{-}
\pi^{-})=(9.3^{+3.7}_{-2.8}\pm 1.9 \pm 2.4)\times 10^{-5}$ with
$8.7\sigma$ and $3.8\sigma$ significance, respectively. If we use
the theoretical calculations $BR(\Xi^{0}_{c}\to \Xi^{-}
\pi^{+})=1.3\%$ and $BR(\Xi^{+}_{c}\to \Xi^{0} \pi^{+})=3.9\%$
\cite{CT_PRD48} and the data $BR(\Xi^{+}_{c}\to \Xi^{0}
\pi^{+})/BR(\Xi^{+}_{c}\to \Xi^{-}\pi^{+}\pi^{+})=0.55\pm 0.16$
\cite{PDG04}, we can get $B^{+}\to \bar \Xi^{0}_{c}
\Lambda^{+}_{c}\approx 4.8 \times 10^{-3}$ and $B^{0}\to \bar
\Xi^{-}_{c} \Lambda^{+}_{c}\approx 1.2\times 10^{-3}$ \cite{CCT}.
Based on these results, one can find that there definitely appear
two puzzles: (I) in terms of the BR of $B^{0}\to \bar
\Sigma^{--}_{c} p \pi^{+}$, one can immediately find that due to
phase space suppression, the BR of $B^{+}\to \Lambda^{+}_{c}
\Lambda^{-}_{c} K^{+}$ is $O(10^{-6})$ which is two orders of
magnitude smaller than the observation; (II) It is expected that
$BR(B\to \bar\Xi_{c} \Lambda^{+}_{c})\sim BR(B^{0}\to\bar
\Lambda^{-}_{c} p )$ in theoretical calculations \cite{QCDth}, but
the reality chooses $BR(B\to \bar\Xi_{c} \Lambda^{+}_{c}) >>
BR(B^{0}\to\bar \Lambda^{-}_{c} p )$. In this paper, we are going to
investigate the possible mechanism to solve the unexpected large BRs
in doubly charmed baryonic B decays.

\begin{figure}[htbp]
\includegraphics*[width=5.5in]{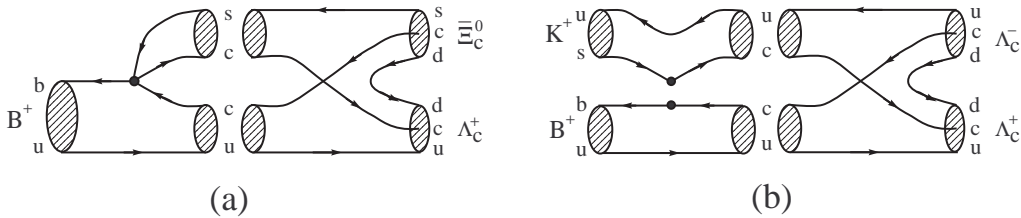}  \caption{ The flavor diagrams for
(a) $B^+\to \bar \Xi^{0}_{c} \Lambda^{+}_{c}$ and (b) $B^{+}\to
\Lambda^{+}_{c} \Lambda^{-}_{c}  K^+$  decays, in which the former
is arisen from $B^{+}\to \bar D^0 D^{+}_{s}$ while the latter is via
$B^{+}\to
 \bar D^{0} D^{0}K^{+}$. The dotted symbols represent the weak
interactions.} \label{fig:flavor}
\end{figure}

It has been noticed that final state interactions (FSIs) may play an
important role on the BRs, CP asymmetries (CPAs) and polarizations
of vector mesons in two-body charmless and charmed mesonic B decays
\cite{HCS_PRD71}. According to our previous discussions, the
inconsistency in two charmed baryons production results from
ordinary estimations and QCD calculations. That is, it should exist
a new mechanism to enhance the BRs for the decays $B\to \bar\Xi_{c}
\Lambda^{+}_{c}$ and $B\to \Lambda^{+}_{c} \Lambda^{-}_{c} K$.
In Ref.~\cite{CCT}, two different mechanisms  have been introduced,
in which the
two-body and three-body decays are governed by the
$\sigma$, $\pi^{0}$ and $\pi^{\pm}$ meson exchanges and
charmonium-like resonance $X_{c\bar{c}}$, respectively.
However, according to the
results of BELLE \cite{belle1}, there is no evidence for the
existence of charmonium-like resonance.
In addition, since the doubly charmed baryonic B decays are found
in two-body and three-body decays simultaneously, we speculate that
they are induced by the similar mechanisms.
Hence, we consider that there exist
other mechanisms which dictate the production of doubly charmed
baryons.
We speculate that FSIs play an essential role on the suffering puzzles.
Furthermore, we propose that the FSIs are arisen from the inelastic
scatterings $\bar D^{0} D^{+}_{s}\to \bar \Xi_{c} \Lambda^{+}_{c}$
and $\bar D^{0} D^{0}  \to \Lambda^{+}_{c} \Lambda^{-}_{c}$ and lead
to $B\to \bar D D^{+}_{s}\to  \bar \Xi_{c} \Lambda^{+}_{c}$ and
$B\to \bar D^{0} D^{0}  K\to\Lambda^{+}_{c} \Lambda^{-}_{c}K$. The
illustrated flavor diagrams are displayed in Fig.~\ref{fig:flavor}.
For convenience, we will use $B\to \bar \Xi_{c}
\Lambda^{+}_{c\,[\bar D D^{+}_{s}]}$ and $B\to \Lambda^{+}_{c}
\Lambda^{-}_{c\, [ \bar D^{0}D^{0}]}K$ instead of the decaying
chains. Since charged $B$ decays have better significance, in the
following analysis we will concentrate on charged modes.

\begin{figure}[htbp]
\includegraphics*[width=3.5in]{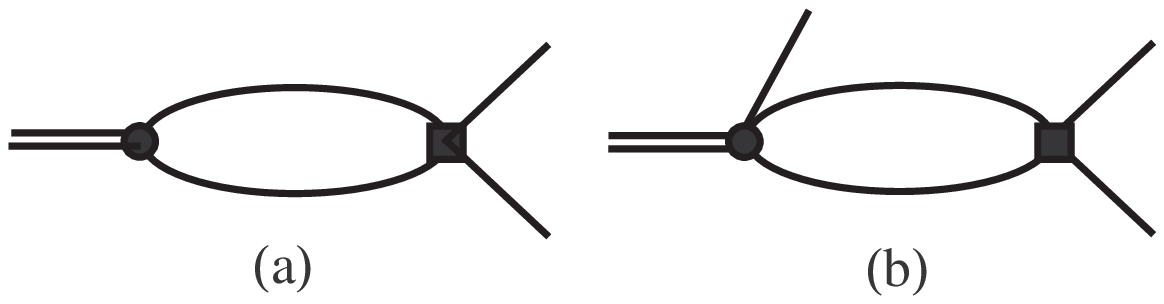}  \caption{The effective diagrams for (a) two-body and (b) three-body
doubly charmed baryonic decays, where double-line denotes the $B$
meson and squared (dotted) symbol stands for the strong (weak)
interactions.}
 \label{fig:effective}
\end{figure}

To be more clear for the calculations and
the effective interactions induced by one-loop effects, we replace
Fig.~\ref{fig:flavor} of flavor diagrams with
Fig.~\ref{fig:effective} of effective diagrams, where (a) and (b)
are for two-body and three-body decays, respectively, and the
squared (dotted) symbol stands for the strong (weak) interactions.
Thus, according to Fig.~\ref{fig:effective}(a),
the decay amplitude for $B\to \bar \Xi_{c} \Lambda^{+}_{c}$ could be
expressed as
\begin{eqnarray}
{\cal M}(B\to \bar\Xi_{c} \Lambda^{+}_{c})&=& \bar{u}_{\Lambda_{c}}
\int \frac{d^4q}{(2\pi)^4} \frac{g_{DD_{s}{\cal B}_{c} {\cal
B}_{c}}(m^2_{B})}{q^2-m^2_{D}+i\varepsilon}
\frac{g_{BDD_{s}}(m^2_B)}{(p_B-q)^2-m^2_{D_s}+i\varepsilon}
v_{\Xi_{c}}, \label{eq:2body}
\end{eqnarray}
where $u(v)$ expresses the Dirac spinor field of baryon
(antibaryon), $\varepsilon$ is used for removing singularities,
$m_{D_{(s)}}$ is the mass of $D_{(s)}$. $g_{BDD_{s}}(m^2_B)$ and
$g_{DD_{s}{\cal B}_{c} {\cal B}_{c}}(m^2_{B})$ denote the effective
weak coupling for $B-D-D_{s}$ interaction and effective strong
coupling for $D-D_{s}-{\cal B}_{c}-{\cal B}_{c}$ interaction with
${\cal B}_{c}$ being a charmed baryon, respectively. Since $D$ and
$D_s$ are the same in the $SU(3)$ flavor symmetry except the carrying
isospins are different, we will assume that the effective strong
interactions are the same for $D$ and $D_s$ except the effects from
the isospin of system. Hence, we will use $g_{DD{\cal B}_{c} {\cal
B}_{c}}$ instead of $g_{DD_{s}{\cal B}_{c} {\cal B}_{c}}$.
Similarly, the same assumption is also applied to charmed baryons.
We note that for two-body decays, because the rest frame of charmed
baryon-pair is just the rest frame of $B$ meson, we have set the
invariant mass of charmed baryon-pair, denoted by $\sqrt{k^2}$, to
be $m_{B}$ the mass of $B$ meson. Therefore, the values of effective
couplings are taken at the $m_{B}$ scale. However, for three-body
decays, the scale should be chosen at $\sqrt{k^2}$ and is a
variable. Although the momentum variable in the loop integration in
principle has no limit,
the main contribution will arise  when the on-shell condition for
the intermediate states $D$ and $D_s$ is satisfied, i.e. we can
regard the production of doubly charmed baryons as the process $B\to
\bar \Xi_{c} \Lambda^{+}_{c\,[\bar D D^{+}_{s}]}$, as illustrated by
Fig.~\ref{fig:flavor}(a). Hence, in terms of narrow width
approximation, $1/[(s-M^2)^2+M^2\Gamma^2_{M}]\approx \pi
\delta(s-M^2)/M\Gamma_{M}$, the integration of Eq.~(\ref{eq:2body})
could be simplified as
\begin{eqnarray}
I_{D_s}(p^2_{B})=\int \frac{d^4q}{(2\pi)^4}
 \frac{1}{q^2-m^2_{D}} \frac{1}{(p_B-q)^2-m^2_{D_s}}\approx
 -\frac{|\vec{p}_{D_{s}}|}{8\pi \sqrt{p^2_{B}}}\label{eq:IDs}
\end{eqnarray}
with $\vec{p}_{D_{s}}$ being the spatial momentum of $D_{s}$ meson.
Since $g_{BDD_{s}}$ is associated with the decays $B\to \bar D
D^{+}_{s}$, we can easily find its relationship to the BRs of $B\to
\bar DD^{+}_{s}$ by
\begin{eqnarray}
BR(B\to \bar{D}D^{+}_{s})=\tau_{B} \frac{|\vec{p}_{D_s}|}{8\pi
m^2_B} |g_{BDD_{s}}(m^2_{B})|^2. \label{eq:brDDs}
\end{eqnarray}
By combining Eqs.~(\ref{eq:2body}), (\ref{eq:IDs}) and
(\ref{eq:brDDs}), we find that the BRs for $B\to \bar\Xi_{c}
\Lambda^{+}_{c}$ decays are related to those for $B\to
\bar{D}D^{+}_{s}$ by
\begin{eqnarray}
BR(B\to \bar\Xi_{c}
\Lambda^{+}_{c})=\frac{|\vec{p}_{\Lambda_c}||\vec{p}_{D_s}|}{32\pi^2}\left(1-
\frac{(m_{\Xi_{c}}+m_{\Lambda_c})^2}{m^2_{B}} \right) |g_{DD{\cal
B}_{c} {\cal B}_{c}}(m^2_{B})|^2 BR(B\to
\bar{D}D^{+}_{s}),\label{eq:brxiclambc}
\end{eqnarray}
where $m_{\Lambda_c(\Xi_{c})}$ is the mass of $\Lambda_{c}(\Xi_{c})$
baryon and $\vec{p}_{\Lambda_c}$ is the spatial momentum of
$\Lambda_{c}$.

 From Eq.~(\ref{eq:brxiclambc}), we clearly see that $BR(B\to
\bar\Xi_{c} \Lambda^{+}_{c})$ can give a constraint on the parameter
$g_{DD{\cal B}_{c} {\cal B}_{c}}(m^2_{B})$.
However, since the constraint is only suitable at the
$m_{B}$ scale,
it will not help us to understand
three-body
decays such as $B\to\Lambda^{+}_{c} \Lambda^{-}_{c} K $ in which the
involving invariant mass of the $\Lambda^{+}_{c}\Lambda^{-}_{c}$ system
is below the $m_{B}$ scale and the value varies between
$2m_{\Lambda_{c}}$ and $m_{B}-m_{K}$.
Moreover, the effects of FSIs should be strongly related to the
momenta of final state particles, so that the contributions of
energetic light baryons are much less important than those of slow
heavy baryons. In sum, for giving a suitable effective strong
coupling for $g_{DD{\cal B}_{c} {\cal B}_{c}}(k^2)$ which could be
applied to various values of invariant mass, we will model the
effective coupling in Lorentz covariant form. In the following, we
show our way to determine the form of effective coupling. If we
regard the scattering $DD\to {\cal B}_c {\cal B}_{c}$ as a t-channel
process in which the intermediate particle is a doubly-charmed
baryon, the scattering amplitude could be approximately described by
\begin{eqnarray*}
\frac{\tilde g^2(k^2)}{m^2_{D}} \bar u_{{\cal B}_{c}} \gamma_{5}
\not{p}_{D} \frac{1}{\not{p}_{D}-\not{ p}_{{\cal B}_{c}}-m_{{\cal
B}_{cc}}} \gamma_{5} \not{p}_{D} v_{{\cal B}_{c}} \approx \frac{\tilde
g^2(k^2)}{m_{{\cal B}_{cc}}} \bar u_{{\cal B}_{c}}v_{{\cal B}_{c}} ,
\end{eqnarray*}
where the fields of D meson have been neglected, $\tilde g(k^2)$
expresses the coupling $D-{\cal B}_{c}-{\cal B}_{cc}$, $m_{{\cal
B}_{cc}}$ denotes the mass of exchanged doubly-charmed baryon and
the small contributions from $p_D- p_{{\cal B}_{c}}$ are neglected.
Comparing to charmless baryonic decays, we speculate that the
dominance of FSIs is due to the particles in the doubly charmed
baryonic decays carrying lower spatial momenta. In order to display
the momentum-dependent strong coupling, we further parametrize the
coupling to be $\tilde g^2(k^2)=g^2_c m^4_{{\cal B}_{cc}}/(|p_{{\cal
B}_{c1}}-p_{{\cal B}_{c2}}|^4+\tilde m^4)$ where $p_{{\cal
B}_{c1(2)}}$ denote the four momenta of charmed baryons, $g_c$ is a
dimensionless coupling constant and $\tilde m$ is the effective
mass, which is used to remove the singularity when $p_{{\cal
B}_{c1}}=p_{{\cal B}_{c2}}$. Accordingly, we model the effective
coupling for the inelastic scattering $DD\to {\cal B}_c {\cal
B}_{c}$ to be
\begin{eqnarray}
g_{DD{\cal B}_{c} {\cal B}_{c}}(k^2)=g^2_{c} \frac{m^3_{{\cal
B}_{cc}}}{ |p_{{\cal B}_{c1}}-p_{{\cal B}_{c2}}|^4+\tilde m^4}.
\label{eq:modeling}
\end{eqnarray}
Since the $m_{{\cal B}_{cc}}$ could be estimated by the masses of
constituent quarks, the undetermined parameters actually are $g_{c}$
and $\tilde m$.

Similar to the decays $B\to \bar \Xi_{c} \Lambda^{+}_{c\, [\bar
DD^{+}_{s}]}$, we find that the production of two charmed baryons in
three-body $B$ decays could be arisen from the the inelastic
scattering such as $B\to \Lambda^{+}_{c} \Lambda^{-}_{c\,[D^{0} \bar
D^{0}]} K  $. According to the Fig.~\ref{fig:effective}(b), the
decay amplitude for $B\to \Lambda^{+}_{c} \Lambda^{-}_{c} K$ could
be written as
\begin{eqnarray}
{\cal M}(B\to \Lambda^{+}_{c} \Lambda^{-}_{c} K)&=&
\bar{u}_{\Lambda_{c}} \int \frac{d^4q}{(2\pi)^4} \frac{g_{DD{\cal
B}_{c} {\cal B}_{c}}(k^2)}{q^2-m^2_{D}+i\varepsilon}
\frac{g_{BKDD}(k^2,\cos\theta)}{(p_B-q)^2-m^2_{D}+i\varepsilon}
v_{\Lambda_{c}},\nonumber \\
&=& C_{I}\, g_{DD{\cal B}_{c} {\cal B}_{c}}(k^2)\,
g_{BKDD}(k^2,\cos\theta)I_D(k^2)
\bar{u}_{\Lambda_{c}}v_{\Lambda_{c}}, \label{eq:3body}
\end{eqnarray}
where $C_{I}$ denotes the factor from the isospin wave function of
system, taking to be $1/\sqrt{2}$ here, and
$g_{BKDD}(k^2,\cos\theta)$ represents the decay amplitude for $B\to
\bar D^0 D^{0}  K$ decays. As known that the decay amplitude is
Lorentz invariant, for convenience, we will set the working frame in
the rest frame of charmed baryon-pair. Hence, $k^2$ stands for the
invariant mass of $\Lambda^{+}_{c} \Lambda^{-}_{c}$ and is a
variable. Angle $\theta$ is the polar angle of $\Lambda^{+}_{c}$
with respect to the momentum direction of $K$ meson. Since $B\to
\bar D^{0} D^0 K$ are color-allowed processes, it is a good
approximation to assume that the processes are dominated by
factorizable effects. Thus, by employing the factorization
assumption, the decay amplitude for $B\to\bar D^0 D^0  K$ could be
written as
\begin{eqnarray}
g_{BKDD}(k^2,\cos\theta)&=&\frac{G_F}{\sqrt{2}} V_{cb}V^{*}_{cs}
a^{\rm eff}_{1} \langle DK| \bar{c} \gamma^{\mu} s|0\rangle \langle
\bar D| \bar{b} \gamma_{\mu}c |B\rangle \label{eq:coupling}
\end{eqnarray}
where $a^{\rm eff}_{1}$ is the effective Wilson coefficient and the
transition matrix elements are defined as
\begin{eqnarray}
\langle \bar D(p_{\bar D})| \bar{b} \gamma_{\mu} c|B(p_{B})\rangle
&=& \left[(p_B+p_{\bar D})_{\mu} -
\frac{m^2_B-m^2_D}{q^2}q_{\mu}\right]F_{1}(q^2)+\frac{m^2_{B}-m^2_{D}}{q^2}q_{\mu}
F_{0}(q^2), \nonumber \\
\langle D(p_D)K(p_K)|\bar{c}  \gamma_{\mu} s|0\rangle &=&
\left[(p_D-p_K)_{\mu}-\frac{m^2_D-m^2_K}{Q^2}Q_{\mu}
\right]F^{DK}_{1}(Q^2) \nonumber \\
&& + \frac{m^2_D-m^2_{K}}{Q^2}Q_{\mu} F^{DK}_{0}(Q^2)
\label{eq:formfactors}
\end{eqnarray}
with $q=p_B-p_{\bar D}$ and $Q=p_D+p_K$. In order to obtain the
information of $g_{BKDD}$ on $k^2$ and $\theta$, we have to analyze
the decays $B\to \bar D^0 D^0 K$ themselves. Since the production of
doubly charmed mesons is similar to that of doubly charmed baryons,
in the following, $k^2$ and $\theta$ will be regarded as the
invariant mass of $\bar D^0D^0$ and $\Lambda^{+}_{c}\Lambda^{-}_{c}$
systems and the relative angle between $\vec{p}_{D^0,\Lambda^{+}_c}$
and $\vec{p}_{K}$, respectively.  For deriving the decay rates as a
function of invariant mass $k^2$ and angle $\theta$, the coordinates
in the $k^2$ rest frame are chosen as follows:
$p_{B(K)}=(E_{B(K)},0,0,|\vec{p}|)$ with $E_{B(K)}=(m^2_B\pm k^2
-m^2_K)/2\sqrt{k^2}$ and $|\vec{p}|=\sqrt{E^2_{K}-m^2_{K}}$ and
$p_{X(\bar X)}=(\sqrt{k^2}/2, \pm |\vec{p}_{X}|\sin\theta,0,
\pm|\vec{p}_{D}|\cos\theta)$ with $|\vec{p}_{X}|=\sqrt{k^2}/2\cdot
\sqrt{1-(2m_X)^2/k^2}$ in which $X$ could be $D^0$ meson or
$\Lambda^+_{c}$ baryon. Based on these coordinates, it is obvious
that $q^2$ and $Q^2$ are functions of $k^2$ and $\theta$. This is
the reason why we set the coupling $g_{BKDD}$ as a function of $k^2$
and $\cos\theta$. Hence, by including the phase space for three-body
decay and using the chosen coordinates, the differential decay rate
for $B\to \bar D^0 D^{0} K$ is expressed by
\begin{eqnarray}
{d\Gamma(B\to \bar D^0 D^{0} K)\over dk^2 d\cos\theta}&=&
\frac{|\vec{p}^{\prime}_K|}{2^8\pi^3m^2_{B}}|g_{BKDD}(k^2,\cos\theta)|^2
\sqrt{1-\frac{(2m_{D})^2}{k^2}} \label{eq:diffDD}
\end{eqnarray}
with $|\vec{p}^{\prime}_{K}|=\sqrt{ E^{\prime 2}_{K}-m^2_{K}}$ and $
E^{\prime}_{K}=(m^2_{B}-k^2+m^2_{K})/2m_{B}$. In terms of
Eq.~(\ref{eq:3body}), similarly, we can also obtain the differential
decay rate for $B\to \Lambda^{+}_{c} \Lambda^{-}_{c} K$ decays as
\begin{eqnarray}
{d\Gamma(B\to \Lambda^{+}_{c} \Lambda^{-}_{c} K) \over dk^2
d\cos\theta} &=& \frac{|\vec{p}^{\prime}_{K}|}{2^7 \pi^3 m^2_{B}}
|I_D(k^2)|^2\left(1-\frac{(2m_{\Lambda_{c}})^2}{k^2} \right)^{3/2}
\nonumber \\
&&\times |g_{BKDD}(k^2,\cos\theta)|^2 |C_{I}\, g_{DD{\cal B}_{c}
{\cal B}_{c}}(k^2)|^2\;. \label{eq:diffLamLam}
\end{eqnarray}
 From Eqs.~(\ref{eq:coupling}),
(\ref{eq:diffDD}) and (\ref{eq:diffLamLam}), we immediately know
that if we can determine $g_{BKDD}(k^2,\cos\theta)$ from $B\to
 \bar D^0 D^{0} K$ and $g_{DD{\cal B}_{c} {\cal B}_{c}}(k^2)$ from $B\to
\bar\Xi_{c}\Lambda^{+}_{c}$, then in principle we can make
predictions on the BRs of $B\to \Lambda^{+}_{c} \Lambda^{-}_{c}K$.

In order to understand whether we can predict the BRs for three-body
doubly charmed baryonic decays, we have to know how many unknown
parameters can be controlled by theoretical calculations and
parametrizations and constrained by experimental measurements. At
first, we discuss the unknowns of Eq.~(\ref{eq:modeling}) for
$g_{DD{\cal B}_{c} {\cal B}_{c}}$. As mentioned before, $m_{{\cal
B}_{cc}}$ could be estimated by the masses of constitute quarks. If
we chose $m_c=1.6$ GeV and the light quark $m_{q}=0.3$ GeV, then we
get $m_{{\cal B}_{cc}}=3.5$ GeV. The estimated value is close to the
particle $\Xi^{+}_{cc}(3520)$, observed recently by SELEX at FNAL
\cite{BCC}. Therefore, there remain two uncertain parameters $g_{c}$
and $\tilde m$ for effective coupling $g_{DD{\cal B}_{c} {\cal
B}_{c}}$. However, the observation of $B^{+}\to \bar \Xi^{0}_{c}
\Lambda^{+}_{c}$ only can constrain their ratio. Thus, without
further assumption, one unknown will be left. Nevertheless, we still
can set the value of $\tilde m$ by a proper approach. As stated
early, the usage of $\tilde m$ is to avoid the suffering problem
when the spatial momentum of charmed baryon in the $k^2$ rest frame
is approaching vanishment. For two-body decays, we can easily know
that the spatial momentum of $\Lambda^{+}_{c}$ is
$|\vec{p}_{\Lambda^{+}_{c}}|=1.15$ GeV. Therefore, the results are
insensitive to $\tilde m$ if its value is less than $1.5$ GeV.
However, for three-body decays, the available range of spatial
momentum of $\Lambda^{+}_{c}$ is
$|\vec{p}_{\Lambda^{+}_{c}}|=[0,0.71]$ GeV. Needless to say, the
value choice of $\tilde m$ will influence on the BRs of $B\to
\Lambda^{+}_{c} \Lambda^{-}_{c} K$ significantly. To get a balance
in both kinds of decays, the $\tilde m$ can be chosen so that its
value will not affect the two-body decays and still can protect
three-body from overestimation.  Since $|p_{{\cal B}_{c1}}-p_{{\cal
B}_{c1}}| \approx 2 |\vec{p_{\Lambda_{c}}}|$ in both two-body and
three-body decays, we find the better value for $\tilde m$ is around
$1.0$ GeV which is similar to the magnitude of momentum in two-body
decays. Hence, if we accept this value as input, $B^{+}\to \bar \Xi^{0}_{c}
\Lambda^{+}_{c}$ can directly give a bound on $g_{c}$.

Next, we concentrate on the effective coupling of
Eq.~(\ref{eq:coupling}). From the equation, we see that the main
unknowns are from the form factors $F_{1(0)}$ and $F^{DK}_{1(0)}$.
We can use a proper QCD approach, such as perturbative QCD
\cite{KLS_PRD67}, relativistic quark model \cite{MS} and light-front
quark model \cite{LF} etc, to calculate the transition form factors
$F_{1(0)}$. According to the results of light-front quark model, the
explicit expressions can be given by \cite{LF}
\begin{eqnarray}
F_{1}(q^2)&=&\frac{0.67}{(1-1.25q^2/m^2_B+0.39(q^2/m^2_B)^2)}, \ \ \
  F_{0}(q^2) = \frac{0.67}{(1-0.65q^2/m^2_B)}.
\end{eqnarray}
As for the form factors $F^{DK}_{1(0)}$, since there exists no good
method to estimate the time-like form factors, what we can do is as
usual to parametrize them to be
\begin{eqnarray}
F^{DK}_1(Q^2)= \frac{F^{DK}(0)}{1 + a^{DK} Q^2/m_X^2 },\ \ \
F^{DK}_0(Q^2)= \frac{F^{DK}(0)}{1 + Q^2/m_X^2}, \label{eq:fDK}
\end{eqnarray}
where we set $m_{X}=m_{B}-m_{D}$ as the maximum invariant mass of
$DK$ system.
We note that the parametrizations in Eq.~(\ref{eq:fDK})
do not display the behavior predicated by PQCD at the large $Q^2$, i.e.
$F^{DK}\to 1/(Q^2\ln(Q^2/\Lambda^2))$ \cite{BF_PQCD};
nevertheless it could be a good approximation
since the
dominant contributions for $B\to \bar{D}^{0}  D^{0}K$ are close to
the threshold region of the invariant mass in the $DK$ system.
In terms of the
parametrizations, we see that two parameters are needed to be
determined. However, so far only the BRs of $B\to\bar D^0 D^{0} K $
are measured. Hence, we only can determine the allowed ratio of
$a^{DK}$ and $F^{DK}(0)$ by the data of $B\to \bar D^{0} D^{0} K$.
If experiments can provide the information on angular distribution,
we can further fix the remaining unknown. Nevertheless, we can still
obtain some information on $F^{DK}(0)$ by considering the two-body
decay $B^{0}\to D^{-}_{s} K^{+}$. It is known that the decay is
governed by annihilation topology \cite{Chen_PLB560} and the BR is
measured to be $(3.8\pm 1.3)\times 10^{-5}$ \cite{PDG04}. In terms
of factorization assumption, the decay amplitude for $B^{0}\to
D^{-}_{s} K^{+}$ can be expressed by
\begin{eqnarray*}
M(B^{0}\to D^{-}_{s} K^{+})&=& \frac{G_{F}}{\sqrt{2}} V_{cb}
V^{*}_{ud} a^{\rm eff}_{2} f_{B} \langle D^{-}_{s} K^{+}|\bar{s}
\not{p}_{B} c |0\rangle \nonumber \\
&=&  \frac{G_{F}}{\sqrt{2}} V_{cb} V^{*}_{ud} a^{\rm eff}_{2} f_{B}
(m^{2}_{D_s}-m^2_{K})F^{D_s K}_{0}(m^{2}_{B})
\end{eqnarray*}
where $f_{B}$ is the decay constant of $B$ meson, $a^{\rm eff}_{2}$
is the color suppressed effective Wilson coefficient and
nonfactorizable effects have been included, and the time-like form
factors defined in Eq.~(\ref{eq:formfactors}) are used. If we
neglect the differences between $D_s$ and $D$ mesons, we can get the
value of form factor $F^{DK}_{0}$ at $m_{B}$ scale. By taking
$V_{cb}=0.041$, $f_{B}=0.2$ GeV, $m_{D_s(K)}=1.969(0.493)$ GeV,
$G_{F}=1.166\times 10^{-5}\ \rm GeV^{-2}$, $\tau_{B^0}=1.55\times
10^{-12}\ \rm s$ \cite{PDG04}, $a^{\rm eff}_{2}=0.275$
\cite{CY_PRD59}, and $BR(B^{0}\to D^{-}_{s} K^{+})=3.5\times
10^{-5}$, we get $F^{DK}_{0}(m^2_{B})\approx
F^{D_sK}_{0}(m^2_{B})=1.0$. Relating the extracted value to the
$F^{DK}_{0}(Q^2)$ parametrized in Eq.~(\ref{eq:fDK}), we immediately
obtain $F^{DK}(0)=3.41$. If we adopt this value as input, the BRs of
$B\to \bar D^{0} D^{0} K $ can fix the parameter $a^{DK}$.
Consequently, by including proper measurements, we can determine the
introduced parameters with errors from experiments.

\begin{figure}[htbp]
\includegraphics*[width=2.5in]{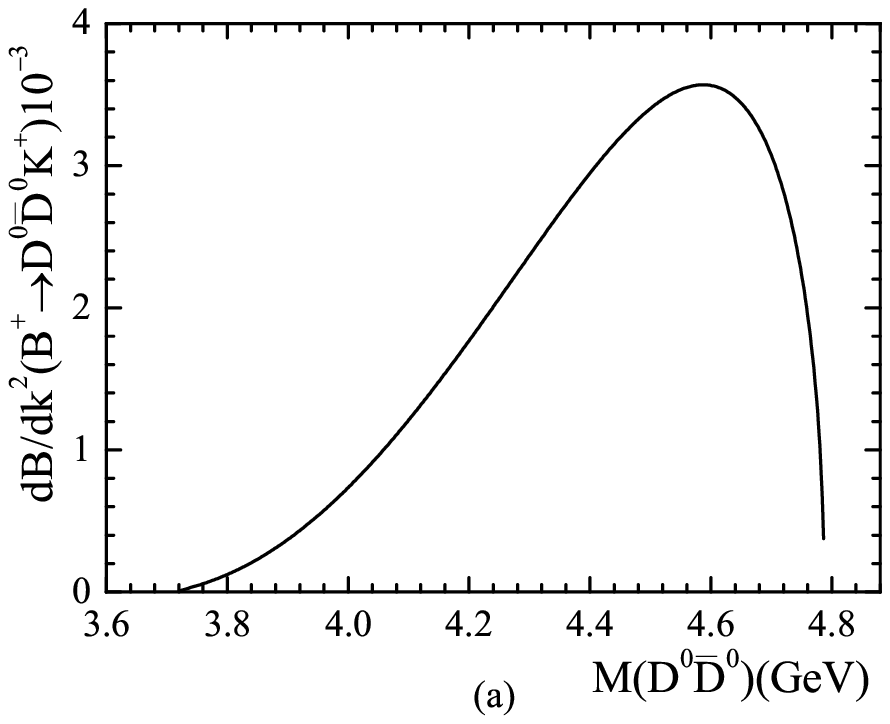}\hspace{1cm}\includegraphics*[width=2.5in]{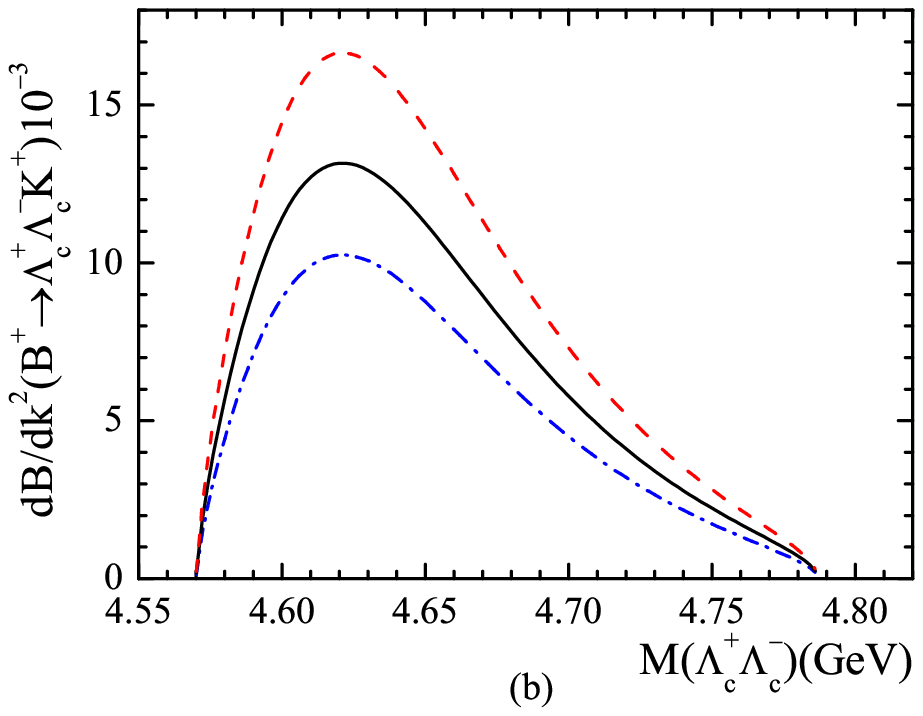}
\caption{The differential BRs for (a) $B^{+}\to\bar D^{0} D^{0}
 K^{+}$ and (b) $B^{+}\to \Lambda^{+}_{c} \Lambda^{-}_{c}
K^{+}$ as a function of invariant mass $M(D^0 \bar D^{0})$ and
$M(\Lambda^{+}_{c}\Lambda^{-}_{c})$, respectively, where we have set
$F^{DK}(0)=3.41$, $a^{DK}=1.0$, $m_{{\cal B}_{cc}}=3.5$ GeV and
$\tilde m=1.0$ GeV. The dash-dotted, sold and dashed lines represent
the contributions when $g_{c}=(3.1,\, 3.3,\, 3.5)$ and their
corresponding BRs for $B^{+}\to\Lambda^{+}_{c} \Lambda^{-}_{c} K^+ $
are $(6.6,\, 8.5,\, 10.9)\times 10^{-4}$, respectively.}
 \label{fig:diff1}
\end{figure}

\begin{figure}[htbp]
\includegraphics*[width=2.0in]{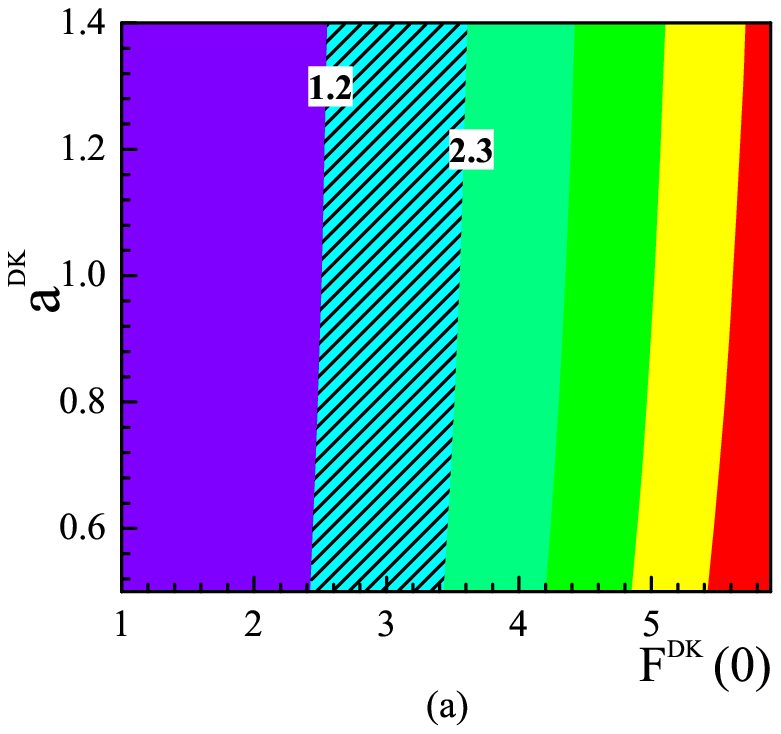}\hspace{0.2cm}
\includegraphics*[width=2.0in]{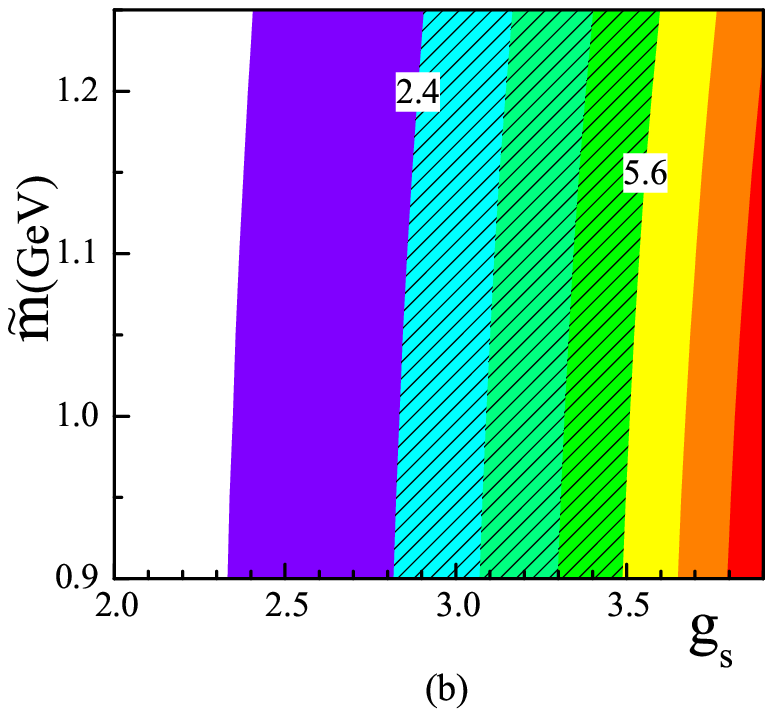}
\hspace{0.18cm} \includegraphics*[width=2.1 in]{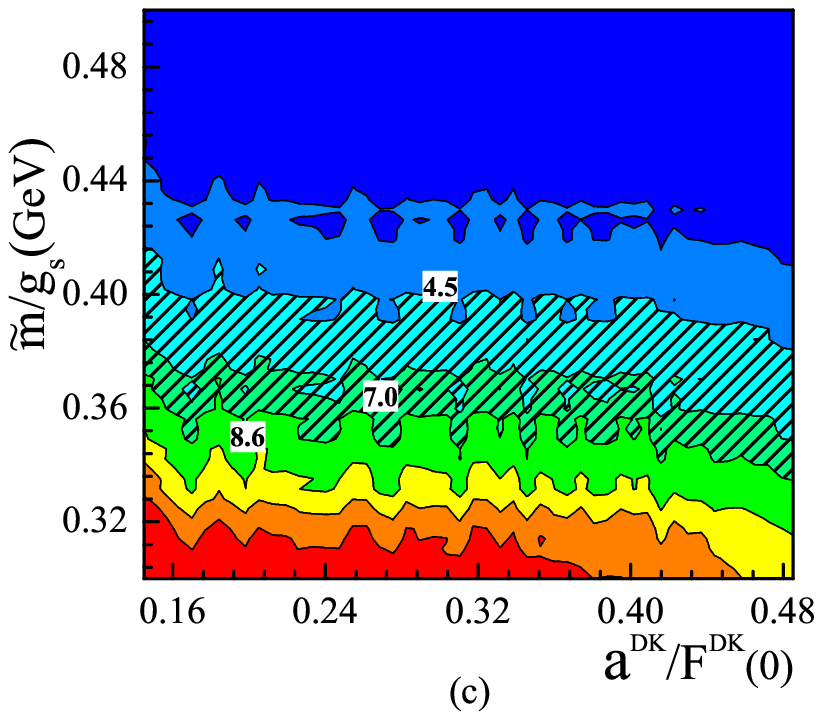} \caption{The
contour plots for  the correlations (a) between $(F^{DK}(0),\,
a^{DK})$ and $BR(B^{+}\to\bar D^{0} D^{0}K^{+} )$ and (b) between
$(g_{c},\,\tilde m)$ and $BR(B^{+}\to \bar\Xi^{0}_{c}
\Lambda^{+}_{c})$. The values (in units of $10^{-3}$) in the figures
(a) and (b) denote the low and upper bounds of current data on BRs.
(c) the possible BR for $B^{+}\to  \Lambda^{+}_{c}
\Lambda^{-}_{c}K^{+}$ when the involving parameters are satisfied
with the bounds of figures (a) and (b), where the values (in units
of $10^{-4}$) shown in the figure come from the data within
$1\sigma$ statistical and systematic errors. }
 \label{fig:constraint}
\end{figure}

In order to calculate the numerical values, besides the values taken
before, we set other inputs as follows: $m_{\Xi^{0}}=2.471$ GeV,
$m_{\Lambda_{c}}=2.285$ GeV, $m_{D^0}=1.865$ GeV,
$\tau_{B^+}=1.67\times 10^{-12}\ \rm s$, $BR(B^{+}\to \bar{D}^0
D^{+}_{s})=(1.3 \pm 0.4)\%$, $BR(B^{+}\to \bar D^{0}  D^0K)=(1.9 \pm
0.4)\times 10^{-3}$ \cite{PDG04} and $a^{\rm eff}_{1}=0.986$
\cite{CY_PRD59}. Although in general it is not necessary to regard
$m_{{\cal B}_{cc}}=3.5$ GeV, $\tilde m=1.0$ GeV and $F^{DK}(0)=3.41$
as the inputs, however, it will be interesting to see whether by
using these values, the predicted BRs of $B\to \Lambda^{+}_{c}
\Lambda^{-}_{c} K$ are consistent with the observations. In
addition, we find that actually the BRs of $B\to \bar D^{0} D^{0} K$
are insensitive to the value of $a^{DK}$. That is, in our
parametrizations, it is a good approximation to take $a^{DK}=1.0$.
Accordingly, we have $BR(B^{+}\to \bar D^{0} D^{0} K^{+}) =2.1\times
10^{-3}$ and the differential BR as a function of invariant mass
$M(D^0 \bar D^0)$ for $B^{+}\to \bar D^{0} D^0 K^+$ is presented in
Fig.~\ref{fig:diff1}(a). To demonstrate the correlation between
$g_{c}$ and $BR(B^{+}\to  \Lambda^{+}_{c} \Lambda^{-}_{c} K^{+})$,
we further set $g_{c}=(3.1,\, 3.3\,,3.5)$ so that the $BR(B^{+}\to
\bar\Xi^0_{c} \Lambda^{+}_{c})=(3.4,\, 4.4,\, 5.6 )\times 10^{-3}$
and $BR(B^{+}\to \Lambda^{+}_{c} \Lambda^{-}_{c} K^{+})=(6.6,\,
8.5,\, 10.9)\times 10^{-4}$; then, we display the corresponding
differential BR for $B^{+}\to \Lambda^{+}_{c} \Lambda^{-}_{c} K^+$
as a function of invariant mass $M(\Lambda^{+}_{c} \Lambda^{-}_{c})$
with dash-dotted, solid and dashed lines, respectively, in
Fig.~\ref{fig:diff1}(b).
 From these results, we see clearly that the observations of
$B\to \bar \Xi_{c} \Lambda^{+}_{c}$ and $B\to\Lambda^{+}_{c}
\Lambda^{-}_{c} K $ could be understood by the similar FSIs.
Furthermore, to
illustrate the influence of
the introduced parameters $F^{DK}(0)$, $a^{DK}$, $\tilde m$ and
$g_{c}$, we free these parameters and let the data decide the
allowed ranges. The results are shown in Fig.~\ref{fig:constraint}.
By Fig.~\ref{fig:constraint}(a), we confirm that the BRs of doubly
charmed mesonic decays are not sensitive to $a^{DK}$. Taking
$1\sigma$ error on BR($B^{+}\to \bar D^0 D^0 K^{+}$), the range for
the value of $F^{DK}(0)$ is $(3,\, 3.9)$. With the bound $(2.4,\,
5.3)\times 10^{-3}$ on $BR(B^{+}\to \bar \Xi^{0}_{c} \Lambda^{+}_{c})$
\cite{belle2},
from
Fig.~\ref{fig:constraint},
we  find
that $BR(B^{+}\to \bar \Xi^{0}_{c} \Lambda^{+}_{c})$
is sensitive to $g_{c}$ but insensitive to $\tilde m$
and we also see
that a wide range of $BR(B^{+}\to \Lambda^{+}_{c}
\Lambda^{-}_{c}K^{+})$  is allowed, including the data shown.
 We note that the values of BR (in units of
$10^{-4}$) shown in the figure (c) only contain $1\sigma$
statistical and systemic errors. By the figure, we also find that
the three-body doubly charmed baryonic decays are much more
sensitive to $g_{c}$ and $\tilde m$, in particular $\tilde m$. Clearly,
based on our modeling of Eq.~(\ref{eq:modeling}), the
parametrizations of Eq.~(\ref{eq:fDK}) and the limited information
on data, in general we cannot make concrete predictions on the
decays $B\to\Lambda^{+}_{c} \Lambda^{-}_{c} K $.

\begin{figure}[htbp]
\includegraphics*[width=2.5in]{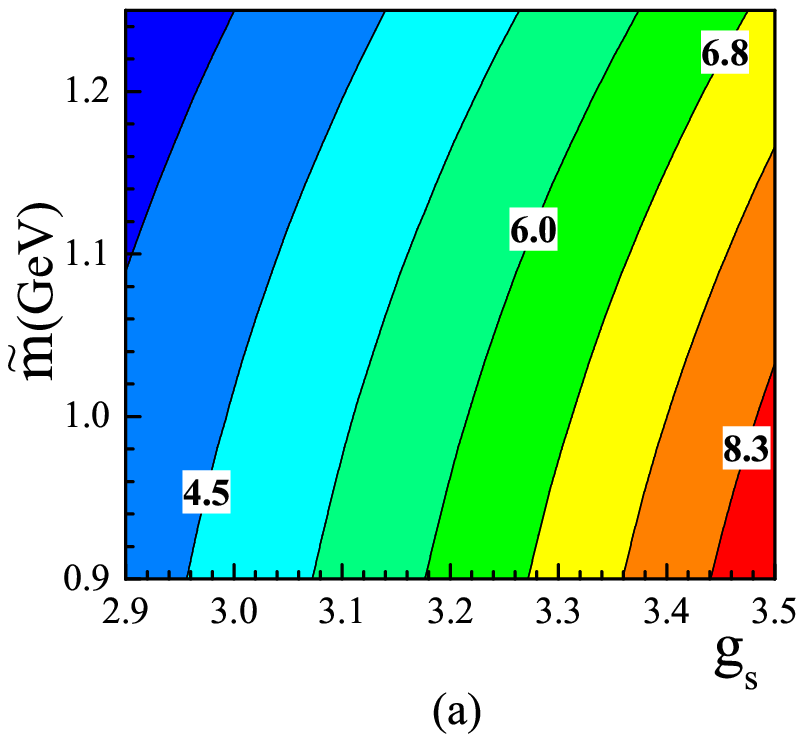}
\hspace{0.2cm} \includegraphics*[width=2.55in]{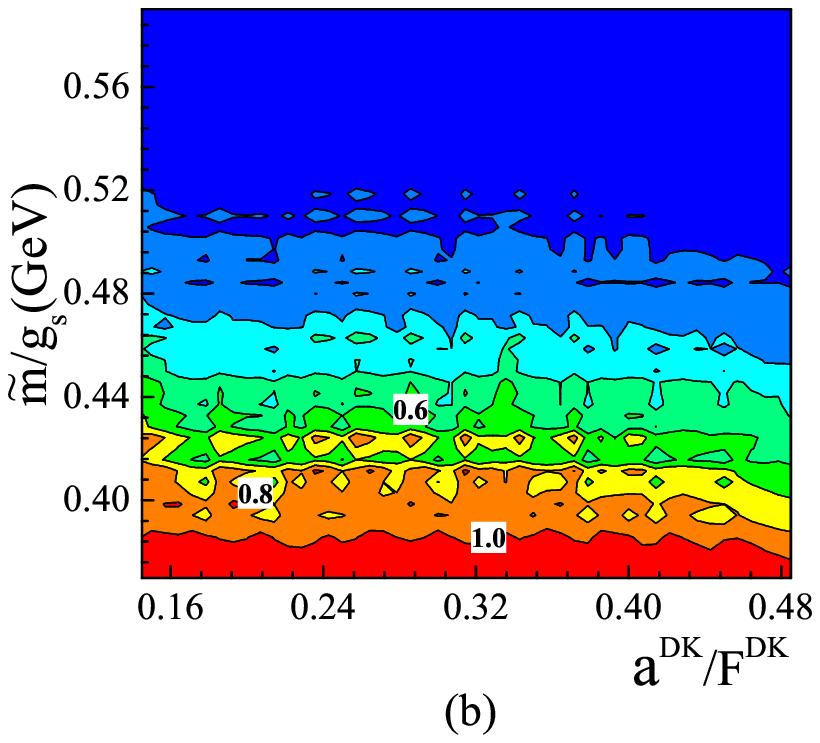}
\caption{The legends of (a) and $(b)$ are the same as
Fig.~\ref{fig:constraint}(b) and (c) but for the decays $B^{+}\to
\bar\Xi^{0}_{c} \Sigma^{+}_{c}$ and  $B^{+}\to \Lambda^{\pm}_{c}
\Sigma^{\mp}_{c}K^{+}$, respectively.}
 \label{fig:predictions}
\end{figure}

Although at present we cannot give precise predictions on doubly
charmed baryonic decays, however, in terms of the proposed mechanisms
we can have some implications on other unobserved processes such as
$B\to \bar \Xi^{0}_{c} \Sigma_{c}$ and $B \to
\Lambda^{\pm}_{c}\Sigma^{\mp}_{c} K$ etc. Since $\Sigma^{+}_{c}$ and
$\Lambda^{+}_{c}$ have the same quark contents except the former is
isospin $I=1$ while the latter is $I=0$,  by utilizing
Eqs.~(\ref{eq:2body}) and (\ref{eq:modeling}) and introducing a
corrected factor $C_{I}=\sqrt{2/3}$ for isospin, we can easily
obtain the BRs for $B\to \bar \Xi^{0}_{c} \Sigma_{c}$. With the
values of $g_{c}$ and $\tilde m$ which are satisfied with the
measurement of $B^{+}\to \bar \Xi^0_{c} \Lambda^{+}_{c}$, the
estimated $BR(B^{+}\to \bar\Xi^{0}_{c} \Lambda^{+}_{c})$ is
presented in Fig.~\ref{fig:predictions}(a) in which we have set
$m_{\Sigma_c}=2.452$ GeV and the values in the figure denote the
available BR in units of $10^{-3}$. It is clear that although the
phase space of the $\bar \Xi^{0}_{c} \Sigma^{+}_{c}$ mode is less
than that of the $\bar\Xi^{0}_{c} \Lambda^{+}_{c}$ mode,  by the
modeling of Eq.~(\ref{eq:modeling}), we predict that  $B^{+}\to \bar
\Xi^{0}_{c} \Sigma^{+}_{c}$ and $B^{+}\to \bar \Xi^0_{c}
\Lambda^{+}_{c}$ have similar BRs, even the former could be larger
than the latter.
Similarly,  in terms of  Eq.~(\ref{eq:diffLamLam})
 with the values of the parameters satisfied with the observation
of $B^{+}\to  \Lambda^{+}_{c} \Lambda^{-}_{c}K^{+}$ within $1\sigma$
error, the estimated $BR(B^{+}\to
\Lambda^{\pm}_{c}\Sigma^{\mp}_{c}K^{+})$ is displayed in
Fig.~\ref{fig:predictions}(b) where the values in the figure are the
available BR in units of $10^{-4}$. From the figure, we see that the
BR of $B^{+}\to  \Lambda^{\pm}_{c}\Sigma^{\mp}_{c}K^{+}$ could be as
large as $10^{-4}$.
 If we extend the FSIs to the
processes associated with the creation of the strange
quark-antiquark pair, i.e. $s\bar s$ pair instead of $d\bar d$ pair
of Fig.~\ref{fig:flavor}(a), and assume that the relevant strong
couplings are the same,
interestingly we obtain $BR(B^{+}\to \bar \Omega_{c}
\Xi^{+}_{c})\approx 5.5\%$. Although $\bar \Omega_{c} \Xi^{+}_{c}$
has much less allowed phase space, due to the enhancement of Eq.~(\ref{eq:modeling}),
we see that the BR for
$B^{+}\to \bar\Omega_{c} \Xi^{+}_{c}$ is not suppressed. However,
since we have assumed that the effective couplings of FSIs for the
production of doubly charmed baryons are the same, one cannot take
the result of $B^{+}\to \bar\Omega_{c} \Xi^{+}_{c}$ seriously. After
all, the FSIs for real situations could be different in different
processes. What we have displayed is that our FSIs could enhance the BR
of process which the corresponding phase space is suppressed.

\begin{figure}[htbp]
\includegraphics*[width=4.5in]{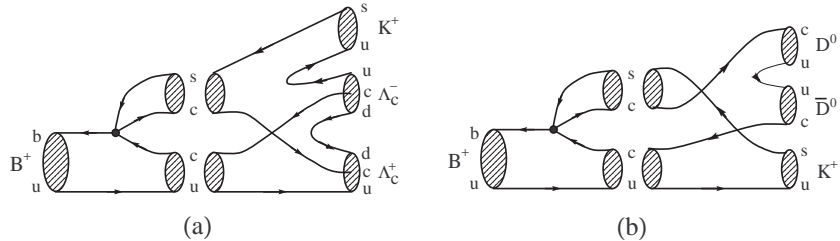}
\caption{  Flavor diagrams (a) for $B^{+}\to \bar{D}^0 D^{+}_{s}\to
\Lambda^{+}_{c} \Lambda^{-}_{c} K^+$ decay and (b) for $B^{+}\to
\bar{D}^0 D^{+}_{s}\to \bar D^{0} D^{0} K^+$ decay.}
 \label{fig:2to3}
\end{figure}

 Finally, we make some remarks on other possible FSIs
for three-body modes, such as $B\to \bar\Xi_{c} \Lambda_{c} \to
\Lambda^{+}_{c} \Lambda^{-}_{c} K$ and $B\to \bar D D^{+}_{s}\to
\Lambda^{+}_{c} \Lambda^{-}_{c} K$. As known that the $BR(B^{+}\to
\Lambda^{+}_{c} \Lambda^{-}_{c} K^{+})$ is only few factors smaller
than $BR(B^{+}\to \bar\Xi^{0}_{c} \Lambda^{+}_{c} )$, if we regard
that the FSIs are inelastic scattering, by the loop and phase space
suppressions, we expect that the contributions through $B\to
\bar\Xi_{c} \Lambda_{c} \to \Lambda^{+}_{c} \Lambda^{-}_{c} K$
should be small. As for the decay $B^{+}\to \bar{D}^0 D^{+}_{s}\to
\Lambda^{+}_{c} \Lambda^{-}_{c} K^{+}$ shown in
Fig.~\ref{fig:2to3}(a), if its contribution is significant, one can
speculate that the same FSIs could also arise from the doubly charmed
mesonic decays such as $B^{+}\to \bar{D}^{0} D^{0} K^{+}$, shown in
Fig.~\ref{fig:2to3}(b). From Fig.~\ref{fig:2to3}(a) and (b), we
conjecture that the former should be smaller than the latter due to
one more quark pair being produced. Since we have assumed that the
dominant effects for $B^{+}\to \bar{D}^{0} D^{0} K^{+}$ are
factorizable contributions, described by Eq.~(\ref{eq:coupling}),
if the decay chain $B^{+}\to \bar D^{0} D^{+}_{s}\to \Lambda^{+}_{c}
\Lambda^{-}_{c} K^{+}$ is not negligible, in our analysis it could be
regarded as a subleading effect.

In summary, we have studied two-body and three-body doubly charmed
baryonic $B$ decays. We find that the observed processes could be
produced by FSIs $ B^{+}\to  \bar \Xi_{c} \Lambda^{+}_{c\,[\bar D
D^{+}_{s} ]}$ and $B\to \Lambda^{+}_{c} \Lambda^{-}_{c\, [\bar
D^{0}D^{0} ]}K$. In terms of proper modeling for the effective
strong coupling and factorization assumption, we find that with the
constrained values of the parameters, the estimated results could be
compatible with the current data. In addition, we also show the
implications of FSIs on $B^{+}\to \bar \Xi^{0}_{c} \Sigma^{+}_{c}$ and
$B^{+}\to \Lambda^{\pm}_{c}\Sigma^{\mp}_{c}K^{+}$ and get that their
BRs could be as large as $BR(B^{+}\to \bar\Xi^{0}_{c}
\Lambda^{+}_{c})$ and $10^{-4}$, respectively. In addition,
extending the FSIs to the processes associated with the creation of
$s\bar s$ pair,
$BR(B^{+}\to \bar \Omega_{c} \Xi^{+}_{c})$ at percent level is achievable. \\

{\bf Acknowledgments}\\

The author would like to thank  Prof. Chao-Qiang Geng and Dr.
Shang-Yuu Tsai for useful discussions. This work is supported in
part by the National Science Council of R.O.C. under Grant
\#s:NSC-94-2112-M-006-009.


\begin{thebibliography}{99}

\bibitem{charmless_baryon1} BELLE Collaboratiuon,
K. Abe {\it et al.}, Phys. Rev. Lett. {\bf 88}, 181803 (2002); M.Z.
Wang, {\it et al.}, Phys. Rev. Lett. {\bf 90}, 201802 (2003); Y.J.
Lee {\it et al.}, Phys. Rev. Lett. {\bf 93}, 211801 (2004).

\bibitem{charmless_baryon2} BABAR Collaboration, B. Aubert {\it et
al.}, Phys. Rev. D{\bf 69}, 091503(R) (2004); BELLE Collaboration,
M.C. Chang {\it et al.}, Phys. Rev. D{\bf 71}, 072007 (2005).


\bibitem{charm_baryon1} BELLE Collaboration, N. Gabyshev {\it et
al.}, Phys. Rev. D{\bf 66}, 091102(R) (2002).

\bibitem{charm_baryon2} BELLE Collaboration, N. Gabyshev {\it et
al.}, Phys. Rev. Lett. {\bf 90}, 12802 (2003); N. Gabyshev {\it et
al.}, arXiv:hep-ex/0409005.

\bibitem{HS}W.S. Hou ands A. Soni, Phys. Rev. Lett. {\bf 86}, 4247 (2001).

\bibitem{CCT} H.Y. Cheng, C.K. Chua and S.Y. Tsai,
arXiv:hep-ph/0512335.

\bibitem{CKM} N. Cabibbo, Phys. Rev. Lett. {\bf 10}, 531 (1963); M. Kobayashi and T. Maskawa, Prog. Theor. Phys.
{\bf 49}, 652 (1973).

\bibitem{Cheng} H.Y. Cheng, arXiv:hep-ph/0603003.

\bibitem{belle1} BELLE Collaboration, N. Gabyshev {\it et al.},
arXiv:hep-ex/0508015.

\bibitem{belle2} BELLE Collaboration, R. Chistov {\it et al.},
arxiv:hep-ex/0510074.

\bibitem{CT_PRD48}H.Y. Cheng and B. Tseng, Phys. Rev. D{\bf 48},
4188 (1993).

\bibitem{PDG04}Particle Data Group, S. Eidelman {\it et al.}, Phys.
Letters B{\bf 592}, 1 (2004).

\bibitem{QCDth}V. Chernyak and I. Zhitnitsky, Nucl. Phys. B{\bf
345}, 137 (1990); P. Ball and H.G. Dosch, Z. Phys. C{\bf 51}, 445
(1991).

\bibitem{HCS_PRD71} H.Y. Cheng, C.K. Chua and A. Soni, Phys. Rev. D{\bf 71}, 014030
(2005).


\bibitem{BCC} SELEX Collaboration, A. Ocherashvili {\it et al.}, Phys. Lett. B{\bf 628}, 18
(2005).

\bibitem{KLS_PRD67}T. Kurimoto, H.N Li and A.I. Sanda, Phys. Rev. D{\bf 67}, 054028
(2003).


\bibitem{MS} D. Melikhov and B. Stech, Phys. Rev. D{\bf 62}, 014006 (2000).

\bibitem{LF} H.Y. Cheng, C.K. Chua and C.W. Hwang, Phys. Rev. D{\bf 69},
074025 (2004).

\bibitem{CY_PRD66} H.Y. Cheng and K.C. Yang, Phys. Rev. D{\bf 66}, 054015
(2002).

\bibitem{BF_PQCD}S.J. Brodsky and G.R. Farrar, Phys. Rev. D{\bf 11}, 1309 (1975).

\bibitem{Chen_PLB560}C.H. Chen, Phys. Lett. B{\bf 560}, 178 (2003).



\bibitem{CY_PRD59} H.Y. Cheng and K.C. Yang, Phys. Rev. D{\bf 59}, 092004
(1999).

\end{thebibliography}
\end{document}